\def\preprint{1}
\ifnum\preprint=9
    \documentclass[11pt,draft,a4paper]{article} 
\fi 
\RequirePackage{etex} 
\def\preprint{1}
\def\masterthesis{0}
\def\bigfont{0} 
\def\cryptology{0} 
\def\sigconf{0} 
\ifnum\sigconf=0
\fi
\def\llncs{0} 
\def\lipics{0} 
\def\quantumjournal{0}

\def\draft{0}
\def\anonymous{0}
\def\shownomenclature{0} 
\def\smallbib{1} 
\def\toc{0} 
\ifnum
    \ifnum\preprint=1 1\else\ifnum\cryptology=1 1\else 0\fi\fi
=1 
    \ifnum\draft=1
       \documentclass[11pt,draft,a4paper]{article}
       \pdfoutput=1
    \else
        \documentclass[11pt,final,a4paper]{article}
        \pdfoutput=1
    \fi
    \usepackage{lmodern}
\fi
\ifnum\masterthesis=1
    \documentclass[11pt,final,a4paper,titlepage]{article}
    \usepackage{lmodern}
\fi
\ifnum\sigconf=1
        \documentclass[sigconf,final]{acmart}
\fi
\ifnum\bigfont=1
    \documentclass[final,17pt]{extarticle} \usepackage{fullpage}
\fi
\ifnum\llncs=1
    \documentclass[runningheads]{llncs}
\fi
\ifnum\lipics=1
    \PassOptionsToPackage{draft=false}{hyperref}

\ifnum\anonymous=0
        \ifnum\draft=0
            \documentclass[a4paper,UKenglish,cleveref, autoref, thm-restate]{lipics-v2021}
        \else
            \documentclass[a4paper,UKenglish,cleveref, autoref, thm-restate, draft]{lipics-v2021}
        \fi
        
    \fi
    \ifnum\anonymous=1
        \ifnum\draft=0
            \documentclass[a4paper,UKenglish,cleveref, autoref, thm-restate, anonymous]{lipics-v2021}
        \else
            \documentclass[a4paper,UKenglish,cleveref, autoref, thm-restate, anonymous, draft]{lipics-v2021}
        \fi
        
    \fi
\title{Template for Papers} 

\author{Or Sattath}{Computer Science Department, Ben-Gurion University of the Negev, Israel}{sattath@bgu.ac.il}{https://orcid.org/0000-0001-7567-3822}{}

\author{Shai Wyborski}{School of Computer Science and Engineering, The Hebrew University of Jerusalem, Israel \and Computer Science Department, Ben-Gurion University of the Negev, Israel}{shai.wyborski@mail.huji.ac.il}{https://orcid.org/0000-0001-6847-5668}{}

\authorrunning{O. Sattath and S. Wyborski} 

\Copyright{Jane Open Access and Joan R. Public} 

\ccsdesc[500]{Hardware~Quantum communication and cryptography}

\keywords{\textcolor{red}{Choose valid keywords}} 

\category{} 

\relatedversion{} 

\acknowledgements{We would like to thank Andrew Miller for valuable discussions and, in particular for pointing out the existence of quantum canaries, and Ori Newman and Elichai Turkel for their invaluable feedback and discussion. This work was supported by the Israel Science Foundation (ISF) grant No. 682/18 and 2137/19, and by the Cyber Security Research Center at Ben-Gurion University. 
\includegraphics[width=40px]{figs/logo-erc}  \includegraphics[width=40px]{figs/logo-eu}}

\EventEditors{John Q. Open and Joan R. Access}
\EventNoEds{2}
\EventLongTitle{Advances in Financial Technologies 2024}
\EventShortTitle{AFT'24}
\EventAcronym{AFT}
\EventYear{2024}
\EventDate{September 23--25, 2024}
\EventLocation{Vienna, Austria}
\EventLogo{}
\SeriesVolume{}
\ArticleNo{}

\fi
\ifnum\quantumjournal=1
  \documentclass[a4paper,onecolumn,11pt 
  ]
  {quantumarticle}
  \pdfoutput=1

\fi

\ifnum\preprint=7 \documentclass{dummy} \fi 

\ifnum\toc=1\ifnum\lipics=0
    \usepackage[nottoc]{tocbibind} 
\fi\fi
\ifnum\lipics=0\ifnum\sigconf=0\usepackage{authblk}\fi\fi 
\ifnum\sigconf=0
    \usepackage{silence} 
    \usepackage{amssymb} 
\fi
\ifnum\lipics=0 
    \usepackage[modulo]{lineno}
\fi
\usepackage{float}
\ifnum\llncs=1

\fi \usepackage{amsthm}

\usepackage{amsfonts,latexsym,color,paralist,url,ifdraft,mathrsfs,thm-restate,comment,bbold,booktabs,ifthen,wrapfig}

\usepackage[utf8]{inputenc} \usepackage[autostyle=false, style=english]{csquotes} \MakeOuterQuote{"}

\usepackage[T1]{fontenc}
\ifnum\sigconf=0
    \ifnum \shownomenclature=1
        \usepackage[refpage,prefix]{nomencl}
         
        \makenomenclature
        
  \makeatletter
        \def\thenomenclature{%
          \providecommand*{\listofnlsname}{\nomname}%

              \section{\nomname}
              \label{sec:nomenclature}    
              \if@intoc\addcontentsline{toc}{section}{\nomname}\fi%

          \nompreamble
          \if@nomentbl
            \let\itemOrig=\item
            \def\item{\gdef\item{\\}}%
            \expandafter\longtable\expandafter{\@nomtableformat}
          \else
            \list{}{%
              \labelwidth\nom@tempdim
              \leftmargin\labelwidth
              \advance\leftmargin\labelsep
              \itemsep\nomitemsep
              \let\makelabel\nomlabel}%
          \fi
        }
        \makeatother 
    \fi
    \ifnum\lipics=0    
        \usepackage[draft=false,pageanchor]{hyperref}
        \usepackage[final]{graphicx}
    \fi
    \usepackage{doi}
\fi

\ifdraft{\usepackage{showlabels}}{} 

\usepackage [ lambda,advantage , operators , sets , adversary , landau , probability , notions , logic ,ff, mm,primitives , events , complexity , asymptotics , keys]{cryptocode}

\ifdraft{\newcommand{\authnote}[3]{{\color{#3} {\bf  #1:} #2}}}{\newcommand{\authnote}[3]{}}

\newcommand{\ket}[1]{\vert #1 \rangle}

\ifdraft{\linenumbers}{}

\usepackage{algorithm,algorithmicx,algpseudocode}
\usepackage[normalem]{ulem}

\usepackage{tikz}
\usetikzlibrary{arrows,chains,matrix,positioning,scopes}
\makeatletter
\tikzset{join/.code=\tikzset{after node path={%
\ifx\tikzchainprevious\pgfutil@empty\else(\tikzchainprevious)%
edge[every join]#1(\tikzchaincurrent)\fi}}}
\makeatother
\tikzset{>=stealth',every on chain/.append style={join},
         every join/.style={->}}
\tikzstyle{labeled}=[execute at begin node=$\scriptstyle,
   execute at end node=$]
\usepackage[capitalise]{cleveref} 
\crefname{Game}{Game}{Games}

\usepackage{autonum}
\ifthenelse{
\(\equal{\cryptology}{1}\OR
\(\equal{\preprint}{1} \OR \equal{\masterthesis}{1}\) \OR \(\equal{\bigfont}{1} \OR \equal{\quantumjournal}{1} \)
\) 
}
{
    \newtheorem{theorem}{Theorem}

    \newtheorem{claim}{Claim}
    \newtheorem{fact}{Fact}
    \newtheorem*{theorem*}{Theorem}
    \newtheorem*{lemma*}{Lemma}
    \newtheorem*{corollary*}{Corollary}
    \newtheorem*{proposition*}{Proposition}
    \newtheorem*{claim*}{Claim}
    \theoremstyle{definition}
    \newtheorem{openproblem}{Open Problem}
    \theoremstyle{remark}
    \newtheorem{remark}[theorem]{Remark}

    \theoremstyle{plain}
    
}{}  
\ifnum\sigconf=1

  \newtheorem*{theorem*}{Theorem}
  \newtheorem*{lemma*}{Lemma}
  \newtheorem*{corollary*}{Corollary}
  \newtheorem*{proposition*}{Proposition}
  \newtheorem*{claim*}{Claim}
  \theoremstyle{definition}

\fi

\theoremstyle{plain}
\newcounter{Game} 
\expandafter\let\expandafter\savedflalignstar\csname flalign*\endcsname
\expandafter\let\expandafter\savedendflalignstar\csname endflalign*\endcsname
\AtBeginDocument{%
  \expandafter\let\csname flalign*\endcsname\savedflalignstar
  \expandafter\let\csname endflalign*\endcsname\savedendflalignstar
}

\newcommand{\ab}[1]{}  

\ifnum\masterthesis=1
    \usepackage{setspace}
    \usepackage{fancyhdr}
    \renewcommand{\headheight}{14pt}
    \newcommand{\longunderline}{\underline{\ \ \ \ \ \ \ \ \ \ \ \ \ \ \ \ \ \ \ \ }\ }
    
    \newenvironment{changemargin}[2]{%
        \begin{list}{}{%
        \setlength{\topsep}{0pt}%
        \setlength{\leftmargin}{#1}%
        \setlength{\rightmargin}{#2}%
        \setlength{\listparindent}{\parindent}%
        \setlength{\itemindent}{\parindent}%
        \setlength{\parsep}{\parskip}%
        }%
    \item[]}{\end{list}}
    
    \newcommand{\logoscale}{0.4}
\fi

\usepackage{xspace}

\DeclareMathAlphabet{\mathpzc}{OT1}{pzc}{m}{it}

\usepackage{booktabs}
\usepackage{longtable}
\usepackage{pgfplotstable}
\pgfplotsset{compat=1.18} 
\usepackage{array}
\usepackage{colortbl}
\begin{document}
\ifnum\masterthesis=0 
    \title{51\% Attack via Difficulty Increase with a Small Quantum Miner
    %
    }
\fi

\ifnum\anonymous=0
    \ifnum\preprint=1
        \author[1]{Bolton Bailey}
        \author[2]{Or Sattath}
        \affil[1]{Department of Computer Science, University of Illinois Urbana-Champaign}
        \affil[2]{Department of Computer Science, Ben-Gurion University of the Negev}
    \fi
    \ifnum\cryptology=1
        \author{XXX}
        \affil{Department of Computer Science, Ben Gurion University of the Negev, Beersheba, Israel\\
        XXX@post.bgu.ac.il}
        \author{Or Sattath}
        \affil{Department of Computer Science, Ben Gurion University of the Negev, Beersheba, Israel\\
        sattath@bgu.ac.il}
    \fi
    \ifnum\sigconf=1
        \author[1]{XXX}
        \author[1]{Or Sattath}
        \affil[1]{Computer Science Department, Ben-Gurion University of the Negev}
    \fi
    \ifnum\llncs=1
        %
        \author{Or Sattath\inst{1}\orcidID{0000-0001-7567-3822} \and
        Second Author\inst{2,3}\orcidID{1111-2222-3333-4444} \and
        Third Author\inst{3}\orcidID{2222--3333-4444-5555}}
        \authorrunning{O. Sattath}
        %
        \institute{Department of Computer Science, Ben Gurion University of the Negev, Beersheba, Israel 
        \email{sattath@bgu.ac.il}\\
        \url{http://www.springer.com/gp/computer-science/lncs} \and
        ABC Institute, Rupert-Karls-University Heidelberg, Heidelberg, Germany\\
        \email{\{abc,lncs\}@uni-heidelberg.de}}
    \fi
    \ifnum\quantumjournal=1
        \author[1]{Andrea Coladangelo}
        \affil[1]{Computing and Mathematical Sciences,	
        Caltech}
        \author[2]{Or Sattath}
        \orcid{0000-0001-7567-3822}
        \email{sattath@bgu.ac.il}

        \affil[2]{Computer Science Department, Ben-Gurion University of the Negev}
        
    \fi
\else
    \ifnum\lipics=0
        \author{}
    \fi
\fi

\ifnum\lipics=1
\keywords{Quantum Mining, Quantum attack on Proof-of-Work}
\fi

\ifthenelse{\equal{\masterthesis}{0} \AND \equal{\sigconf}{0}}{
    \maketitle
} 
{}        
\ifnum\masterthesis=1
    \begin{titlepage}
        \centering
        { Ben-Gurion University of the Negev}
        
        {The Faculty of Natural Sciences}
        
        {\small The Department of Computer Science}
        
        \vspace{2cm}
        
        \includegraphics[scale=\logoscale]{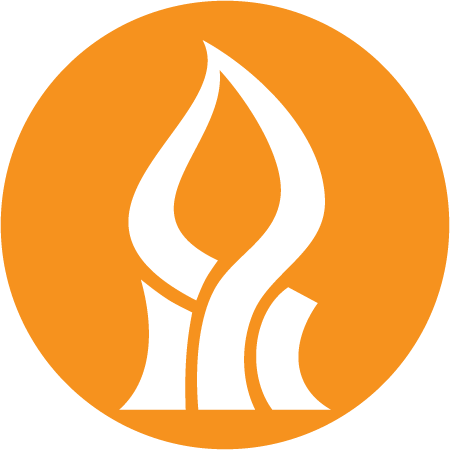}
        
        \vspace{2cm}
        
        {\Large \bfseries Template of Thesis}
        
        \vspace{2cm}
        
        {\small Thesis submitted in partial fulfillment of the requirements for the Master of Sciences degree}
        
        \vspace{1cm}
        
        {\bfseries Name of Student}
        
        {Under the supervision of Dr. Or Sattath}
        
        \vspace{2cm}
        
        \today
    \end{titlepage}
    
    \begin{titlepage}
        \centering
        { Ben-Gurion University of the Negev}
        
        {The Faculty of Natural Sciences}
        
        {\small The Department of Computer Science}
        
        \vspace{2cm}
        
        {\Large \bfseries Template of Thesis}
        
        \vspace{2cm}
        
        {\small Thesis submitted in partial fulfillment of the requirements for the Master of Sciences degree}
        
        \vspace{1cm}
        
        {\bfseries Name of Student}
        
        {Under the supervision of Dr. Or Sattath}
        
        \vspace{1cm}
        
        {\small Signature of student: \longunderline Date: \longunderline}
        
        \vspace{0.5cm}
        
        {\small Signature of supervisor: \longunderline Date: \longunderline}
        
        \vspace{0.5cm}
        
        \begin{changemargin}{-1.5cm}{-1.5cm}
        \centering
            {\small Signature of chairperson of the committee for graduate studies: \longunderline Date: \longunderline}
        \end{changemargin}
        \vspace{2cm}
        
        \today
    \end{titlepage}

    \pagenumbering{roman}
    \begin{center}
        {\large \bfseries Title of Thesis}
        
        \vspace{0.5cm}
        
        {\bfseries Name of Student}
        
        \vspace{0.5cm}
        
        Thesis submitted in partial fulfillment of the requirements for the Master of Sciences degree
        
        \vspace{0.25cm}
        
        {Ben-Gurion University of the Negev}
        
        \vspace{0.25cm}
        
        \today
        
        \vspace{2cm}
        
        {\bfseries Abstract}
        
        \vspace{0.5cm}

        Thesis abstarct goes thesis
    \end{center}
    \pagebreak
    \pagenumbering{arabic}
    \subsection*{Acknowledgments}
    \setcounter{tocdepth}{3}
    
    \pagebreak
    \tableofcontents
    \listoffigures 
    \listoftables 
    \pagebreak
\else 
    \ifnum\sigconf=0
        \begin{abstract}
    \fi
\fi
\ifnum\sigconf=0
    \ifnum\masterthesis=0

    We present a strategy for a single quantum miner with relatively low hashing power, with the same ramifications as a 51\% attack.
    Bitcoin nodes consider the chain with the highest cumulative proof-of-work to be the valid chain. A quantum miner can manipulate the block timestamps to multiply the difficulty by $c$. 
    The fork-choice rule counts every block with increased difficulty with weight $c$. 
    By using Grover's algorithm, it is only $\sqrt c$ harder for the quantum miner to mine such blocks. 
    By picking a high enough $c$, the single quantum miner can create a competing chain with fewer blocks, but more cumulative proof-of-work. 
    The time required is $O(\frac{1}{r^2})$ epochs, where
    $r$ is the fraction of the speed at which the quantum miner can produce blocks with Grover's algorithm, compared to the honest network.

    Most proof-of-work cryptocurrencies, including Bitcoin, are vulnerable to our attack. However, it will likely be impossible to execute in forthcoming years, as it requires an extremely fast and fault-tolerant quantum computer. 
  
    \fi
\fi 
\ifnum\masterthesis=0
    \ifnum\sigconf=0
        \end{abstract}
    \fi
\fi

\ifnum\sigconf=1
    \begin{abstract}
        Abstract of SigConf goes here.
    \end{abstract}
    \keywords{}
    \maketitle
\fi

\ifnum\toc=1
    \ifnum\llncs=1
        \setcounter{secnumdepth}{3}
        \setcounter{tocdepth}{3}
    \fi
    \tableofcontents 
\fi

\section{Introduction} 
\label{sec:introduction}
In Proof-of-Work (PoW)~\cite{DN92}, a prover can prove to other parties that a predefined computational effort has been spent. Is PoW "compatible" with quantum computing? More specifically, can the prover get \emph{more} than a quadratic advantage due to Grover's algorithm~\cite{Gro96}? Here, we study this question in the context of cryptocurrencies.
It is often stated that a single quantum miner does not pose a security risk to Bitcoin's security.
This position is supported by the formal analysis made in \cite{CGK+23}, which shows some desired properties (namely, common prefix and chain quality) are satisfied in a simplified model called the Bitcoin backbone protocol in a network consisting of a single quantum miner and additional (honest) classical miners. 

This work shows that the contrary is true by exploiting an aspect that Ref.~\cite{CGK+23} does not account for.

A single quantum miner, with relatively little hash power, can execute an attack in the presence of honest classical miners with the same ramifications as a 51\% attack. 
Specifically, the attacker can double-spend transactions and censor others, thus breaking Bitcoin's security, and can receive $1-\epsilon$ of the block rewards for any $\epsilon>0$.

The attack relies on the ``chain work'' consensus mechanism in Bitcoin and many other PoW based cryptocurrencies: In case of competing forks, Bitcoin nodes follow the branch which has the most accumulated PoW, see~\cite[Section 4]{Nak08} and \cite[pp. 239--240]{Ant17}. In short forks, where different branches have the same difficulty, this coincides with the longest chain rule; but this may not be so in long forks, which span over more than one epoch and can have different difficulty adjustments: the sum of difficulties in the shorter chain may be higher than that in the longer chain. This consensus mechanism thwarts certain classical attacks (we discuss one such attack in more detail in \cref{sec:longest_chain}), but it leaves the protocol vulnerable to quantum attacks. 

\begin{remark}
Our results do not contradict those in ~\cite{CGK+23}: as mentioned, their results hold in a \emph{simplified} model. In their model, there are no difficulty adjustments; therefore, the longest chain rule and the chain with the most cumulative PoW coincide. In practice, a difficulty adjustment mechanism must be introduced; otherwise, due to fluctuations in the miners' hash power, the rate at which blocks are created can be either higher or slower than desired. An earlier version of their work lists difficulty adjustments as an important future direction: "The first generalization to consider is the Bitcoin backbone protocol with variable difficulty"~\cite[Section 6]{CGK+19}. As they point out, such a generalization \emph{was} considered in the classical setting~\cite{GKL17}; our work shows that the quantum generalization does not apply.
\end{remark}

\begin{remark}
    This work studies a setting with a \emph{single} quantum miner. The fully quantum setting, where all miners use quantum hardware, is mostly uncharted territory. Ref.~\cite{Sat20} shows that Bitcoin's tie-breaking rule needs to be adapted. Ref.~\cite{LRS19} analyzes the equilibria strategies for quantum miners in a simplified model. 
\end{remark}

The main advantage quantum miners have over classical miners is that they can use Grover's algorithm. 
We now explain the essential properties of Grover's algorithm for this work at a high level. 

Let $f:\{1,\ldots,n\}\to \{0,1\}$. We call $x$ a marked item if and only if $f(x)=1$, and denote by $k$ is the number of the marked items in $f$. We assume that the function $f$ is modeled as a black box, and every query made to the function is counted. Classically, it can easily be shown that one needs $\theta(\frac{n}{k})$ queries to the function to find a marked item with constant probability. Grover's algorithm gives a quadratic speed-up. That is, the quantum algorithm finds a marked item with high probability using only $\Theta(\sqrt{\frac{n}{k}})$ quantum queries to the function. 
Throughout this work, for the sake of simplicity, we ignore the constant factor in the $\Theta(\cdot)$ notation and the probabilistic nature of Grover's algorithm and treat it as if its running time is $\sqrt{\frac{n}{k}}$ and success probability is $1$.
In the case of mining, the function $f$ is defined as the function that returns $1$ if the hash of a block with a nonce $x$ is below the difficulty, and furthermore, we model the hash function as in the random oracle model.  No further understanding of quantum computing or Grover's algorithm will be needed. The interested reader is referred to Ref.~\cite{NC00} for an in-depth introduction to quantum computing.

The most important insight for the attack is the following: A difficulty increase by a factor of $c$ makes it $c$ times harder for classical miners to find a block, but only $\sqrt c$ harder for quantum miners using Grover's algorithm to do so. 
A block with such increased difficulty is accounted as $c$ blocks with the original difficulty for the accumulated PoW, but is only $\sqrt c$ harder for for the quantum miner to find.
By increasing the difficulty by a factor of $c$, a quantum miner can increase the rate at which they add cumulative work to a chain by a factor of $\sqrt{c}$.

We will describe the main difficulty increase attack in this section and three more variants in \cref{sec:variants}, each improving or addressing a problem in the previous variant. The only downside of these improved variants is that they take longer to execute. 

For concreteness, we use a running example with the same parameters as in Bitcoin: the expected time to create a block by the classical miners is 10 minutes, and difficulty adjustments are made every epoch, which lasts 2016 blocks.
Creating an epoch worth of blocks takes two weeks if a block is created every 10 minutes, and we refer to this period as \emph{epoch-time}.\footnote{
Note that it may take much longer than one epoch-time for the quantum miner to mine one epoch of blocks, especially as the difficulty in that epoch increases.}     
We further assume that the quantum miner can mine a block every $40$ minutes. In other words, by running Grover's algorithm, the miner can create a block at \textcolor{blue}{$r=\frac{1}{4}$} of the speed relative to the classical miners.\footnote{In our running example, we label numbers derived from this choice of $r$ in \textcolor{blue}{blue}.}

\begin{remark}
Note that a quantum miner that can generate a block at a speed ratio $r$ compared to the classical miners might generate less than $r$ of the total revenue from these blocks if they were to mine on top of the honest chain tip, due to stale blocks. Stale blocks are blocks that the quantum miner generates but would eventually not be part of the longest chain due to other blocks being mined simultaneously. This phenomenon has some surprising consequences observed and analyzed by Nerem and Gaur~\cite{NG23}, but will not be relevant to our work: Since the attacker mines a chain independent of the honest network, they are not constrained by the possibility that some of their blocks will be made stale.
    \label{rem:peaceful_mining_lower_revenue}
\end{remark}

\paragraph*{Main attack.}
The attacker's first task is to increase the difficulty by a factor of $\frac{4}{r^2}=\textcolor{blue}{64}$, where $r=\frac{1}{4}$  mentioned above. To achieve this first task, the attacker starts to mine an entire epoch with fake timestamps such that the blocks seem to be created $\frac{4}{r^2}=\textcolor{blue}{64}$ times more often than they should be.
Recall that difficulty adjustments aim to ensure blocks are created at a designated rate: one block is expected every $10$ minutes in Bitcoin. At the end of the epoch, since the timestamps in these blocks appeared to occur every $\textcolor{blue}{\frac{10}{64}}$ minutes, the difficulty in the attacker's chain is increased by a factor of $\textcolor{blue}{64}=\frac{4}{r^2}$, as desired.\footnote{This is not entirely accurate, see p.~\pageref{par:incompatible_difficutly}, and Variant 4 which resolves this issue.} 

The miner's second task is to create another epoch of blocks at the new difficulty.
In more detail, the miner creates one epoch worth of blocks---that is 2016 blocks---with the new difficulty (which is $64$ times harder than the original one), by spacing the timestamps at the usual 10 minutes. The miner publishes these blocks, while ignoring blocks created by the classical miners. 

We claim that after the first epoch with the increased difficulty is published, the quantum miner has a chain with more accumulated PoW than the classical miners. To see this, let us first calculate how long it takes for the quantum miner to achieve the first goal. Recall that it takes $\frac{10}{r}=\textcolor{blue}{40}$ minutes to mine a single block with the original difficulty by the quantum miner, which is $4$ times as much the classical miners take in expectation. So, it will take $\frac{1}{r}=\textcolor{blue}{4}$ epoch-times (\textcolor{blue}{8} weeks) to achieve the first goal by the quantum miner.

The difficulty of the second epoch is $\frac{4}{r^2}=\textcolor{blue}{64}$ times higher, so mining it would take the attacker $\sqrt{\frac{4}{r^2}}=\frac{2}{r}=\textcolor{blue}{8}$ times more than the first epoch; in other words, quantum mining an epoch with the new difficulty will take $\frac{2}{r}\cdot \frac{1}{r}=\frac{2}{r^2}=\textcolor{blue}{32}$ epoch-times.


The quantum miner accumulated $\textcolor{blue}{65}=1+\frac{4}{r^2}\geq \frac{4}{r^2}$ epochs worth of PoW measured in the original difficulty. The time it took is $\textcolor{blue}{36}=\frac{1}{r}+\frac{2}{r^2}\leq \frac{3}{r^2}$ epochs.
So, we see that it took the quantum miner at most $\frac{3}{r^2}$ epoch-times, to generate at least $\frac{4}{r^2}$ epochs worth of PoW. Therefore, the quantum miner's chain has at least $\frac{1}{r^2}$ more cumulative PoW than the expected cumulative PoW by the classical miners, so with high probability, the quantum miner's chain will be the chain with the most accumulated PoW.

This variant allows the quantum miner to double spend a transaction: the quantum miner can mine his chain privately and, while mining, spend a transaction on the classical miners' chain, and post a conflicting transaction on his chain. After this transaction is confirmed (since the quantum miner is mining in private, the transaction will have many confirmations), and the goods are received, the miner can publish his private chain, which will eventually have more PoW, hence causing a long-term reorganization, causing the original transfer to get canceled. 
The details of the Variant 1 attack are depicted in~\cref{tab:variant1}.
\pgfplotstableset{ col sep=comma,every head row/.style={before row=\toprule,after row=\midrule},every last row/.style={after row=\bottomrule}} 
\begin{table}
    \centering
    \resizebox{\linewidth}{!}{\begin{filecontents*}{matlabCode/variant1.csv}
n,difficulty,CPoW,timeToCreate,realTimeWhenCreated,timestamp
1,1,1,4,4,0.015625
2,64,65,32,36,1.015625
\end{filecontents*}
\pgfplotstabletypeset{matlabCode/variant1.csv}
    }

    \caption{The main attack. 
    The column headers represent the following: \textbf{n} is the epoch number, \textbf{difficulty} is the difficulty of that epoch relative to the original difficulty, \textbf{CPoW} is the cumulative Proof-of-Work of the quantum miner, \textbf{timeToCreate} is the time the quantum miner invested in creating this epoch, measured in epoch-time (for example, 1 represents 2 weeks for Bitcoin), \textbf{realTimeWhenCreated} is the real time, measured in epoch-times, when that epoch was created, and \textbf{timestamp} represents the timestamps that appear in the last block in that epoch of the quantum miner, measured in epoch-time. All the other tables have the same format.
    }
    \label{tab:variant1}
\end{table}

The quantum miner can keep mining at the new difficulty, while ignoring the blocks generated by the classical miners, and produce $100\%$ of the blocks in the chain with the most cumulative PoW.

This strategy has several issues that are discussed and addressed in several variants in \cref{sec:variants}. The first issue is that the time-stamps are lagging: the timestamps in the blocks generated by the quantum miner lag behind real time; This issue is addressed in Variant 2. The next issue is related to the incentives. The quantum miner that uses the strategy above creates only 2 epochs worth of blocks, even though the total time of the attack is 36 epochs. In Variant 3, we show an attack where the quantum miner's revenue gets arbitrarily close to the maximal revenue, namely, the revenue generated by an honest classical miner with $100\%$ of the hashing power.
In the variants above, we apply sharp changes to the mining difficulty. Bitcoin limits the increase and decrease factor of the difficulty by a factor of four. This can be easily addressed, by making smaller changes to the difficulty, as demonstrated in Variant 4. The main drawback of these variants compared to the attack above, is that they take longer to perform, though, asymptotically, they all take $O\left(\frac{1}{r^2}\right)$ epoch-times to implement.

\section{Variants}
\label{sec:variants}

\subsection{The Lagging Issue}
Even though the quantum miner can create a chain with higher cumulative PoW, the timestamps in the chain in the variant above are lagging behind the real time.
More specifically, in the example above, one can notice that the timestamp in the second row is 1.02 epoch-times, whereas the real time is 36 epoch-times.
Therefore, the attack can easily be noticed. Furthermore, an attempt to rule out the attack can be made, by changing the fork choice rule: blocks that have timestamps that substantially lag behind the real time can be ignored or alerted by honest nodes. This is the case for the Bitcoin Core client, which will remain in ``initial block download'' state if its chain tip is over 24 hours old, and will alert the user. 

In the 2nd variant, when the quantum attacker's chain is published, the final time-stamp does not lag behind the real time. The modified fork-choice rule would permit the fork of the quantum miner and, therefore, is ineffective for the second variant. 

\paragraph*{Variant 2.}
The goal of this variant is to resolve the lag that is created in the first variant. \cref{tab:variant2} shows the details of the attack. 
The attack proceeds in three phases: After the difficulty is increased in the first phase, as in Variant 1, there is a second phase in which the quantum miner \emph{reduces} the difficulty, to a lower level than the original difficulty, which allows the quantum miner to generate blocks much faster in the third phase. 
Specifically, the difficulty is ultimately reduced to $r^6=\textcolor{blue}{\frac{1}{4^6}}$.\footnote{While it is technically only necessary to reduce the difficulty to below $r^2$ to make up ground on the timestamp gap, our choice of $r^6$ makes the analysis more clean.} This is done by reducing the current difficulty (which is $\frac{4}{r^2})$ by a factor of 8 for $\log_8(\frac{4}{r^8})=\textcolor{blue}{6}$ epochs.
\footnote{Note that the alternative in which the difficulty is decreased in 1 epoch by this factor would take much longer: It would require making an average timestamp increase of $\frac{4}{r^8}$ block times, leading to a timestamp \textit{in the future}. This solves the issue at hand, where the timestamps in the attacker's blocks are noticeably behind the real time, but introduces the \emph{opposite} problem. 

The way to reduce the difficulty by a factor $d$ (up to rounding errors) while inducing the least change in timestamp is to reduce the difficulty for $\ln(d)$ epochs, where the difficulty decreases by a factor $e$ between every two epochs.
}

The time that it takes to mine the hardest epoch is $\frac{2}{r^2}=\textcolor{blue}{32}$ epoch-times, and every consecutive epoch has an easier difficulty by a factor 8, so each consecutive block takes only a fraction of $\frac{1}{\sqrt{8}}$ the time compared to the previous epoch. Overall, it takes $$\frac{2}{r^2}\sum_{i=0}^{\log_8(\frac{4}{r^8})-1} \left(\frac{1}{\sqrt 8}\right)^i\leq \frac{2}{r^2}\sum_{i=0}^\infty \left(\frac{1}{\sqrt 8}\right)^i = \frac{2}{r^2(1-\frac{1}{\sqrt 8})} 
$$
epoch-times for the quantum miner to reduce the difficulty to the easy level.

Recall that the quantum miner can generate each block at the original difficulty every $\frac{1}{r}=\textcolor{blue}{4}$ blocks. 
Due to the decreased difficulty, and Grover's algorithm running time, mining a block at that reduced difficulty takes only $\frac{\sqrt{r^6}}{r}=r^2=\textcolor{blue}{\frac{1}{16}}$ epoch-times. The timestamps are kept at the normal pace (i.e., a block every 10 minutes), so the lag is getting shorter. The quantum miner keeps mining until there is no significant gap between the timestamp and when the block is created. In this example, as seen in \cref{tab:variant2}, the lag is shortened by mining $5$ additional blocks at the easy difficulty level.
As can be seen, there is no significant lag between the timestamp and the time the blocks are published: 53.72 vs 53.02 epoch-times (and can be made arbitrarily small easily).

Recall that the goal of the last phase is to make sure that there is no lag between the timestamp and the real-time in the last epoch of the attack. 
The time it takes for the quantum miner to mine each epoch at the easy difficulty is $\sqrt{r^6} \cdot \frac{1}{r} = r^2$ epoch-times, whereas the timestamp grows by exactly 1 epoch-time each epoch, so the gap between the timestamp and the real time shrinks by $1-r^2$ epoch-times each mined epoch.
The total time that needs to be gapped is at most the total time of the attack in the first two phases, which is upper-bounded by $(\frac{1}{r} + \frac{2}{r^2(1-\frac{1}{\sqrt 8})})$. So, the total time devoted by the quantum miner to mine at the easy difficulty is at most  $(\frac{1}{r} + \frac{2}{r^2(1-\frac{1}{\sqrt 8})})\cdot\frac{r^2}{1-r^2}$ epoch-times. The overall epoch-time of the attack (that is, by adding the time for the previous phase) is 
\[ \left( \frac{1}{r} + \frac{2}{r^2(1-\frac{1}{\sqrt 8})} \right)\cdot \left(1+\frac{r^2}{1-r^2} \right)
\leq \frac{3.57}{r^2},\]
where we used $r \leq \frac{1}{4}$ in the last inequality.

\begin{remark}
    Notice that for attackers with a value of $r$ close to 1, the attack as laid out takes a long time, because we do not reduce the difficulty to a very small value. For this reason it is convenient to assume that an attacker with $r > \frac{1}{4}$ simply carries out the attack as if $r = \frac{1}{4}$, so that they decrease the difficulty sufficiently in the latter stage. This preserves the $O(\frac{1}{r^2})$ asymptotics.
\end{remark}


So, the expected cumulative PoW that the honest network can generate is at most as above. The quantum miner generates $\frac{4}{r^2}$ PoW in the second epoch of the attack alone, so the accumulated PoW on the quantum attacker's chain is greater than the expected accumulated PoW of the classical miners. 

\begin{table}
    \centering
    \resizebox{\linewidth}{!}{\begin{filecontents*}{matlabCode/variant2.csv}
n,difficulty,CPoW,timeToCreate,realTimeWhenCreated,timestamp
1,1,1,4,4,0.015625
2,64,65,32,36,8.015625
3,8,73,11.3137084989848,47.3137084989848,16.015625
4,1,74,4,51.3137084989848,24.015625
5,0.125,74.125,1.4142135623731,52.7279220613579,32.015625
6,0.015625,74.140625,0.5,53.2279220613579,40.015625
7,0.001953125,74.142578125,0.176776695296637,53.4046987566545,48.015625
8,0.000244140625,74.142822265625,0.0625,53.4671987566545,49.015625
9,0.000244140625,74.14306640625,0.0625,53.5296987566545,50.015625
10,0.000244140625,74.143310546875,0.0625,53.5921987566545,51.015625
11,0.000244140625,74.1435546875,0.0625,53.6546987566545,52.015625
12,0.000244140625,74.143798828125,0.0625,53.7171987566545,53.015625
\end{filecontents*}
\pgfplotstabletypeset{matlabCode/variant2.csv}
    }
    
    \caption{The table shows the details of Variant 2. 
    Note that in the strategy, the timestamp in epoch 12 is $53.02$ epoch-times, while the real time lags only slightly behind it $53.72$. This should be compared vis-a-vis the first variant, where the lag was substantially higher. 
    Note that during that 53 epoch-times, the miner has mined only 12 epochs worth of blocks. Therefore, the block subsidy from the quantum mining relative to a miner that mines all the blocks with the original difficulty is only $23\%$; this will be improved in the third variant. }
    \label{tab:variant2}
\end{table}

\subsection{The Revenue Issue}
In Variant 2, the attacker creates only $\textcolor{blue}{12}$ epochs worth of blocks, even though in real time, the attack takes about $\textcolor{blue}{53}$ epoch-times. This means that the revenue during the attack from the block subsidy is only $\textcolor{blue}{\approx 23\%}$ of the total revenue that can be generated from mining at the original speed. Typically, a 51\% attacker would accumulate 100\% of the expected block subsidy for the time period they carry out the attack.

Next, we show a variant in which the revenue from the block subsidy can get arbitrarily close to the total revenue that can be extracted by a classical miner with 100\% of the hashing power.

\paragraph*{Variant 3.}
Increasing the block subsidy revenue is done by adding several epochs in the increased difficulty, and additional epochs in the decreased difficulty, until the timestamp and the time in which the quantum miner's chain is published are close enough. In the example shown in \cref{tab:variant3}, three epochs are mined at the increased difficulty (compared to one epoch in Variant  2), and \textcolor{blue}{72} epochs with the decreased difficulty (compared to five epochs in Variant 2).
The total time in this variant is \textcolor{blue}{121.84} epochs. During it, the quantum miner generates \textcolor{blue}{80} epochs worth of blocks, yielding approximately $\textcolor{blue}{66\%}$ of the block subsidy revenue that the entire network, mining at the original speed, would have generated. This improves upon the $23\%$ of the block subsidy revenue generated in Variant 2. In \cref{sec:intcreased_revenue} we discuss this attack in more generality, where the attack achieves $1-\epsilon$ fraction of the revenue generated honestly, for an arbitrary small $\epsilon$, and takes $O(\frac{1}{\epsilon r^2})$ epochs. 

\begin{table}
    \centering
    \resizebox{\linewidth}{!}{
    \pgfplotstableset{every nth row={13}{after row={\\ & & &\vdots & & \\ \\}}} 
    \begin{filecontents*}{matlabCode/variant3short.csv}
n,difficulty,CPoW,timeToCreate,realTimeWhenCreated,timestamp
1,1,1,4,4,0.015625
2,64,65,32,36,1.015625
3,64,129,32,68,2.015625
4,64,193,32,100,10.015625
5,8,201,11.3137084989848,111.313708498985,18.015625
6,1,202,4,115.313708498985,26.015625
7,0.125,202.125,1.4142135623731,116.727922061358,34.015625
8,0.015625,202.140625,0.5,117.227922061358,42.015625
9,0.001953125,202.142578125,0.176776695296637,117.404698756654,50.015625
10,0.000244140625,202.142822265625,0.0625,117.467198756654,51.015625
11,0.000244140625,202.14306640625,0.0625,117.529698756654,52.015625
12,0.000244140625,202.143310546875,0.0625,117.592198756654,53.015625
13,0.000244140625,202.1435546875,0.0625,117.654698756654,54.015625
14,0.000244140625,202.143798828125,0.0625,117.717198756654,55.015625
78,0.000244140625,202.159423828125,0.0625,121.717198756654,119.015625
79,0.000244140625,202.15966796875,0.0625,121.779698756654,120.015625
80,0.000244140625,202.159912109375,0.0625,121.842198756654,121.015625
\end{filecontents*}
\pgfplotstabletypeset{matlabCode/variant3short.csv}
    }
    \caption{The table shows the details of Variant 3.
    Note that epochs 15--77 were omitted; their difficulty is the same as epoch 14, and the effect is similar to the epochs below and above them. 
    }
    \label{tab:variant3}
\end{table}

\subsection{The Permitted Difficulty Adjustments Issue}

\label{par:incompatible_difficutly}
In some cryptocurrencies, the shifts in the PoW difficulty adjustments are bounded. For example, in Bitcoin, the difficulty between two consecutive epochs cannot%
\footnote{See \url{https://github.com/bitcoin/bitcoin/blob/4b1196a9855dcd188a24f393aa2fa21e2d61f061/src/pow.cpp\#L56}.
} increase or decrease by more than a factor of $4$. 
In all the previous variants, we increased the difficulty dramatically in the first epoch and reduced it by a factor of 8 in several consecutive epochs: all these transitions are incompatible with Bitcoin's difficulty adjustments. This aspect is addressed in the following variant.

\begin{table}[h!t]
    \centering
    \resizebox{\linewidth}{!}{
    \pgfplotstableset{every nth row={30}{after row={\\ & & &\vdots & & \\ \\}}} 
    \begin{filecontents*}{matlabCode/variant4short.csv}
n,difficulty,CPoW,timeToCreate,realTimeWhenCreated,timestamp
1,1,1,4,4,0.5
2,2,3,5.65685424949238,9.65685424949238,1
3,4,7,8,17.6568542494924,1.5
4,8,15,11.3137084989848,28.9705627484771,2
5,16,31,16,44.9705627484771,2.5
6,32,63,22.6274169979695,67.5979797464467,3
7,64,127,32,99.5979797464467,3.5
8,128,255,45.254833995939,144.852813742386,4
9,256,511,64,208.852813742386,5
10,256,767,64,272.852813742386,6
11,256,1023,64,336.852813742386,8
12,128,1151,45.254833995939,382.107647738325,10
13,64,1215,32,414.107647738325,12
14,32,1247,22.6274169979695,436.735064736294,14
15,16,1263,16,452.735064736294,16
16,8,1271,11.3137084989848,464.048773235279,18
17,4,1275,8,472.048773235279,20
18,2,1277,5.65685424949238,477.705627484771,22
19,1,1278,4,481.705627484771,24
20,0.5,1278.5,2.82842712474619,484.534054609518,26
21,0.25,1278.75,2,486.534054609518,28
22,0.125,1278.875,1.4142135623731,487.948268171891,30
23,0.0625,1278.9375,1,488.948268171891,32
24,0.03125,1278.96875,0.707106781186548,489.655374953077,34
25,0.015625,1278.984375,0.5,490.155374953077,36
26,0.0078125,1278.9921875,0.353553390593274,490.508928343671,38
27,0.00390625,1278.99609375,0.25,490.758928343671,40
28,0.001953125,1278.998046875,0.176776695296637,490.935705038967,42
29,0.0009765625,1278.9990234375,0.125,491.060705038967,43
30,0.0009765625,1279,0.125,491.185705038967,44
31,0.0009765625,1279.0009765625,0.125,491.310705038967,45
539,0.0009765625,1279.4970703125,0.125,554.810705038967,553
540,0.0009765625,1279.498046875,0.125,554.935705038967,554
541,0.0009765625,1279.4990234375,0.125,555.060705038967,555
\end{filecontents*}
\pgfplotstabletypeset{matlabCode/variant4short.csv}
    }
    \caption{The table shows the details of Variant 4.
    As can be seen, the difficulty adjustments in this variant are at most $2$ and at least $\frac{1}{2}$, compared to the previous variants, where the increases in difficulty were much higher, and the decrease in difficulty reached a factor of $\frac{1}{8}$; these sharp changes are not in the valid regime of Bitcoin's difficulty adjustments, which is $[\frac{1}{4},4]$. Note that the relative revenue in this variant increased to $\frac{541}{555.06}\approx 97\%$.
    }
    \label{tab:variant4}
\end{table}
\paragraph*{Variant 4.}
As mentioned above, we need to reduce the rate of change in the difficulty adjustments: Instead of increasing the difficulty in the first epoch, we do so by increasing the difficulty in several epochs, where the difficulty is increased each epoch by a factor of 2; 
Similarly, instead of reducing the difficulty by a factor of 8, we reduce it by a factor of 2, which triples the number of epochs needed for that difficulty reduction. 

The attack is shown in \cref{tab:variant4}. Increasing by a factor of 2 in each epoch, it takes 8 epochs to reach the maximum difficulty of 256, and decreasing by a factor of 2 each epoch, it then takes 18 epochs to reach the minimum difficulty of $\approx 9.77 \cdot 10^{-4}$. Ultimately, though, these transitions only require a number of epochs logarithmic in the desired difficulty change, and their effect on the cumulative PoW and timestamp change is dominated by the contributions from the blocks at the top and bottom difficulties, leaving the total running time still at $O(\frac{1}{r^2})$.

\section{Applicability} \label{sec:applicability}

As far as we are aware, in theory, the attack applies to all existing PoW cryptocurrencies such as Bitcoin~\cite{Nak08}, as well as variants thereof, which all need to have some form of difficulty adjustments~\cite{SZ15,PS17}.

The attack described in this work cannot be executed with existing technologies because of two main reasons: (i) The quantum computers have on the order of (only) 100 qubits, and (ii) these quantum computers are too noisy to perform long computations. 
 
Next, we analyze the minimal requirements of a quantum computer to apply this attack. The analysis we make is optimistic in the sense that we underestimate the time analysis of the attack.

Examples of this underestimation are the following: 


\begin{itemize}
    \item We assume that the quantum computer is noiseless. A tighter analysis would account for a poly-logarithmic factor associated with error-correction.
    \item Bitcoin applies SHA-256 \emph{twice} on the header, where we account only for one.
    \item The circuit depth analysis we use is for single-block SHA-256. A Bitcoin block-header is 80 bytes and requires multiple blocks.
    \item When applying Grover's algorithm, the oracle needs to be implemented by a \emph{clean} computation, which involves computing the function, and then uncomputing it. We ignore the uncomputing.
    \item Each Grover's iteration need to apply a reflection about the $\ket{+++\cdots+}$ state, which we ignore.
    \item Our analysis uses the current total hash-rate of the Bitcoin network as the basis for the attack, where the quantum miner would have to use future total hash-rates, which, based on past trends, will likely be higher by the time quantum computers capable of carrying out even a single Grover iteration are developed.
\end{itemize}

As will be demonstrated next, our quantum attack takes centuries to execute even under these relaxations, so it is clearly unrealistic. A tighter and more detailed analysis of the resources needed for quantum mining was done in Ref.~\cite{ABL+18}.

As of February 2024, Bitcoin hash-rate is around 500 exa-hashes per second (EH/S)$=5 \cdot 10^{20}$ H/S.\footnote{See, e.g., \url{https://www.coinwarz.com/mining/bitcoin/hashrate-chart}} 
Since a block is created every 600 seconds, it means that the probability to win a block per hash is $1/(600\cdot 5\cdot 10^{20})= \frac{1}{3\cdot 10^{23}}$. 
To find a block, a miner needs to apply the order of $\sqrt{3\cdot 10^{23}}$ Grover iterations. 
Implementing a SHA-256 evaluation can be done via a circuit with a depth of 1600\footnote{See \url{https://nigelsmart.github.io/MPC-Circuits/}.}. 
So, for a miner to mine a block on a single device, requires a quantum computation with depth in the order of $1600\cdot \sqrt{3\cdot 10^{23}} \lesssim 10^{15}$. 
Suppose a fast, noiseless quantum computer operates at a 10GHz clock speed, slightly surpassing the 9GHz (classical) world record\footnote{See \url{https://www.intel.com/content/www/us/en/newsroom/news/13th-gen-crosses-9-ghz-overclocking-record.html}.} clock speed. Such a quantum computer would find a block every $10^{15}/10^{10}=10^5$ seconds, which is roughly once a day.
This is about $140$ times slower than the existing block rate. So, naively, such a single quantum computer would have $r\approx 0.7\%$. 
Recall that the attack proposed in our work takes the order of $r^{-2}$ epochs $\approx 800$ years, which is clearly unrealistic under the assumptions above. There are several routes in which the attack could become realistic: (i) the clock speed of the quantum computer has to be more than an order of magnitude faster than the fastest commercially available classical computers. (ii) Attack cryptocurrencies with less total hash-rate, or much shorter epoch-time. (iii) Use many quantum computers in parallel in the attack. Note that Grover's algorithm does not lend itself to perfect parallelization as classical unordered search: Running Grover's algorithm on $k$ quantum computers cuts the time only by a $\frac{1}{\sqrt{k}}$ multiplicative factor~\cite[Section IV]{Zal99}.

\section{Conclusion and Open problems}

Our work has demonstrated a variety of attacks on Nakamoto consensus using difficulty adjustment. These attacks are not possible or practical with the quantum computers of today. In the long-term though, they represent a challenge to the thesis that 51\% of network hashpower is necessary to execute a double-spend on Bitcoin. At the least, our work shows that hopes of a security proof for this consensus system that realizes both quantum and difficulty adjustment will not be realized.

We conclude with some potential directions for future work:

\paragraph*{Impossibility results for more general PoW consensus mechanisms} Our work here deals with standard Nakamoto consensus. However, there are a variety of other forms of consensus that use PoW which involve multiple chains or even directed-acyclic graph structures \cite{bagaria2019prism, PS17, SZ15, SLZ16}. While the cited works do not always cover difficulty adjustment, it seems likely that it could be added in some form to most of these other consensus mechanisms. Future work could explore whether there are stumbling blocks akin to the issues covered in Section \ref{sec:variants} that prevent the difficulty increase attack for these mechanisms. In particular, do limitations on the timestamp deltas or bounds on per-epoch difficulty change help prevent the attack in these systems?

\paragraph*{Optimal attacks under different attacker preferences} Malicious actors may want to attack blockchains for a variety of reasons. At a large scale, nation-states may be opposed to blockchains for reasons relating to financial regulation and control. On a smaller scale, there may be criminal organizations who simply want to extract monetary value from the attack, either by double-spending or by using the event of a chain compromise to short the market. The plethora of attack models imply different circumstances as to the resources of the attackers, and their relative preference on the speed and value of the attack. Future work could consider these possibilities in more specificity, and determine the optimal attack parameters for specific goals.

\paragraph*{Mitigations} As we discussed in Section \ref{sec:applicability}, one of the limiting factors on the attack is the size of the circuit used for the hash function. This suggests one way of mitigating an attack of the type we propose---simply making the hash function more difficult to compute, such as repeated hashing. This is unfortunately only a stopgap, though, since Grover's algorithm can ultimately be applied to any PoW system that takes the form of a repeated search problem. Future work might consider if there is some way of structuring the problems that miners solve so that they cannot be sped up by Grover search, see, for example, using Momentum as an alternative to SHA-256 for the purpose of Proof-of-Work, which was suggested by~\cite{ABL+18}. Alternatively, one could consider if there are other ways of structuring the difficulty adjustment and the relative weights assigned to blocks of different difficulty that defeat this type of attack.

\ifnum\lipics=0
    \ifnum\masterthesis=0
        \subsection*{Acknowledgments}
    \fi
    
    \ifnum\anonymous=0
               The authors would like to thank Andrew Miller for discussions on the topic of this paper.

            \BeforeBeginEnvironment{wrapfigure}{\setlength{\intextsep}{0pt}}
            \begin{wrapfigure}{r}{100px}
                \includegraphics[width=100px]{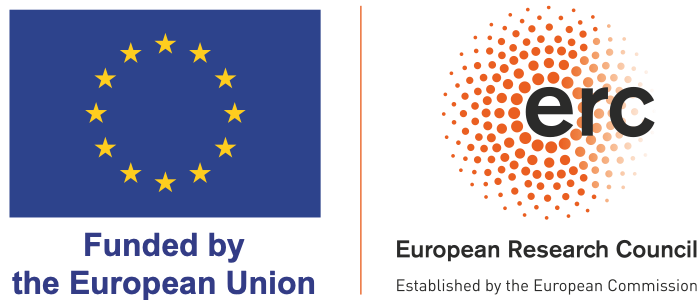}
            \end{wrapfigure}
    This work was funded by the European Union (ERC-2022-COG, ACQUA, 101087742). Views and opinions expressed are however those of the author(s) only and do not necessarily reflect those of the European Union or the European Research Council Executive Agency. Neither the European Union nor the granting authority can be held responsible for them.

            This material is based upon work supported by the National Science Foundation under the Graduate Research Fellowship Program with Grant No. DGE – 1746047.
    \fi
\fi
\ifnum\sigconf=1
    \bibliographystyle{ACM-Reference-Format}
\else
    \ifnum\cryptology=1
        \bibliographystyle{abbrv}
    \else
        \ifnum\lipics=1
            \bibliographystyle{plainurl}
        \else
            \bibliographystyle{alphaabbrurldoieprint}
    \fi
\fi

\ifnum\masterthesis=0
    \ifnum\smallbib=1
        {\footnotesize \bibliography{main} }
    \else 
        \bibliography{main}
    \fi
\fi

\appendix
\ifnum\shownomenclature=1
\printnomenclature[1in]
\fi

\section{Increased revenue}
\label{sec:intcreased_revenue}
Recall that Variant 3 allows a quantum miner to increase the revenue up to $(1-\epsilon) b$, where $b$ is the revenue that an honest miner controlling $100\%$ of the hashing power will generate during that period. \cref{tab:variant3better} demonstrates an attack with $\epsilon=0.05$, i.e., the quantum miner generates $95\%$ of the total revenue. A similar analysis to that in Variant 2 shows that the time of the attack scales as $O(\frac{1}{\epsilon r^2})$ epochs.
\begin{table}
    \centering
    \resizebox{\linewidth}{!}{
    \pgfplotstableset{every nth row={31}{after row={\\ & & &\vdots & & \\ \\}}} 
    \begin{filecontents*}{matlabCode/variant3bettershort.csv}
n,difficulty,CPoW,timeToCreate,realTimeWhenCreated,timestamp
1,1,1,4,4,0.015625
2,64,65,32,36,1.015625
3,64,129,32,68,2.015625
4,64,193,32,100,3.015625
5,64,257,32,132,4.015625
6,64,321,32,164,5.015625
7,64,385,32,196,6.015625
8,64,449,32,228,7.015625
9,64,513,32,260,8.015625
10,64,577,32,292,9.015625
11,64,641,32,324,10.015625
12,64,705,32,356,11.015625
13,64,769,32,388,12.015625
14,64,833,32,420,13.015625
15,64,897,32,452,14.015625
16,64,961,32,484,15.015625
17,64,1025,32,516,16.015625
18,64,1089,32,548,17.015625
19,64,1153,32,580,18.015625
20,64,1217,32,612,19.015625
21,64,1281,32,644,20.015625
22,64,1345,32,676,21.015625
23,64,1409,32,708,22.015625
24,64,1473,32,740,23.015625
25,64,1537,32,772,31.015625
26,8,1545,11.3137084989848,783.313708498985,39.015625
27,1,1546,4,787.313708498985,47.015625
28,0.125,1546.125,1.4142135623731,788.727922061358,55.015625
29,0.015625,1546.140625,0.5,789.227922061358,63.015625
30,0.001953125,1546.142578125,0.176776695296637,789.404698756655,71.015625
31,0.000244140625,1546.14282226562,0.0625,789.467198756655,72.015625
32,0.000244140625,1546.14306640625,0.0625,789.529698756655,73.015625
33,0.000244140625,1546.14331054688,0.0625,789.592198756655,74.015625
794,0.000244140625,1546.3291015625,0.0625,837.154698756655,835.015625
795,0.000244140625,1546.32934570312,0.0625,837.217198756655,836.015625
796,0.000244140625,1546.32958984375,0.0625,837.279698756655,837.015625
\end{filecontents*}
\pgfplotstabletypeset{matlabCode/variant3bettershort.csv}
    }
    \caption{The table shows another variant, where the relative revenue is
    $95\%$, which improves upon the $66\%$ achieved in Variant 3. This variant is closely related to Variant 3, except it has 24 epochs mined at the highest difficulty (compared to only 3 epochs in Variant 3), and additional blocks at the easiest difficulty to reduce the lag.}
    \label{tab:variant3better}
\end{table}

\section{Problems with the longest chain rule}
\label{sec:longest_chain}
Our attack crucially relies on Bitcoin's fork-choice rule, which prefers the chain with the most cumulative PoW. One may consider a natural alternative, in which the fork-choice rule is changed to the longest-chain, ignoring difficulty changes between different epochs. Next, we show an example where a small miner can double-spend and generate the entire revenue from block rewards in this model, assuming the other miners are honest, if the modified fork-choice rule is used.

Consider the following setting. Suppose that the network starts with a genesis block with difficulty 1, and in the next epoch, the difficulty is increased to 1000, and stays with that difficulty for $n$ additional epochs.
A single miner with $\frac{1}{500}$ of the total hashing power, can create a long range fork from the genesis block, in which the difficulty does not increase. To ensure that the attacker's chain does not lag behind the others, in the last epoch, the attacker publishes the timestamps at a slightly faster pace than the one in the first honest epoch. 

Note that the cumulative PoW in the honest chain is $1+1000\cdot n$ whereas the attacker's cumulative work is only $1+n$. Since the last epoch is created slightly faster, the attacker has a longer chain, which would get selected by the longest chain rule. This attack would not work with Bitcoin's existing fork-choice rule that chooses the chain with the most cumulative PoW, since the attacker's chain has only a tiny fraction of cumulative PoW compared to the honest chain.  

\end{document}